\documentclass[aps,prl,preprint,amsmath,amssymb,superscriptaddress,showpacs,showkeys]{revtex4-1}
\usepackage{graphicx}
\usepackage{subcaption}
\usepackage{textcomp}
\usepackage{graphicx}
\usepackage{dcolumn}
\usepackage{bm}
\usepackage[utf8]{inputenc}
\usepackage[english]{babel}
\usepackage{color}
\usepackage{amsmath, amssymb, amsthm, mathrsfs}
\usepackage{hyperref}
\usepackage{natbib}
\usepackage{epsfig}
\usepackage{amsfonts}
\usepackage{xcolor}
\usepackage{braket}
\usepackage[toc,page,title,titletoc,header]{appendix}
\definecolor{olivegreen}{rgb}{0.0, 0.5, 0.0}
\newcommand{\ketbra}[2]{|#1\rangle\langle#2|}

\begin{document}
\title{Spatial characterization of photonic polarization entanglement using an intensified Tpx3Cam fast camera}
\author{Christopher Ianzano}
\affiliation{Department of Physics \& Astronomy, Stony Brook University, Stony Brook, NY 11794-3800, USA}
\author{Peter Svihra}
\affiliation{Department of Physics, Faculty of Nuclear Sciences and Physical Engineering, Czech Technical University,
Prague 115 19, Czech Republic}
\author{Mael Flament}
\affiliation{Department of Physics \& Astronomy, Stony Brook University, Stony Brook, NY 11794-3800, USA}
\author{Andrew Hardy}
\affiliation{Brookhaven National Laboratory, Upton NY 11973, USA}
\affiliation{Department of Physics, University of Toronto, Toronto ON M5S 3H7, Canada}
\author{Guodong Cui}
\affiliation{Department of Physics \& Astronomy, Stony Brook University, Stony Brook, NY 11794-3800, USA}
\author{Andrei Nomerotski}
\affiliation{Brookhaven National Laboratory, Upton NY 11973, USA}
\author{Eden Figueroa}
\affiliation{Department of Physics \& Astronomy, Stony Brook University, Stony Brook, NY 11794-3800, USA}

\begin{abstract}
Scalable technologies to characterize the performance of quantum devices are crucial to creating large quantum networks and quantum processing units. Chief among the resources of quantum information processing is entanglement. Here we describe the full temporal and spatial characterization of polarization-entangled photons produced by Spontaneous Parametric Down Conversions using an intensified high-speed optical camera, Tpx3Cam. This novel technique allows for precise determination of Bell inequality parameters with minimal technical overhead, as well as novel characterization methods of the spatial distribution of entangled quantum information. This could lead to multiple applications in Quantum Information Science, opening new perspectives for the scalability of quantum experiments.
\end{abstract}

\maketitle

\section{Introduction}

\noindent Ever since the original experiments with entangled photons \cite{Kwiat1995}, photonic entanglement has become a remarkable resource in the development of quantum technologies, including entanglement over long-distance for quantum communication \cite{Tittel1998}, entanglement swapping \cite{Pan1998}, teleportation between a photon and an atomic ensemble \cite{Sherson2006}, violation of the CHSH (Clauser-Horne-Shimony-Holt) inequality measured over long distances \cite{Ursin2007,Hensen2015} and entanglement of spin waves among four quantum memories \cite{Choi2010}.\\

\noindent The creation of quantum networks of many such quantum devices in which entanglement is shared among multiple network nodes is the next technological frontier for the successful development of these applications. Easy-to-use, scalable, and compact characterization devices, providing all the information regarding entanglement in near-real-time are fundamental for further large-scale network developments.\\

\noindent Recent developments have shown that spatial characterization of entangled states with single-photon sensitive cameras provides access to a myriad of new possibilities, such as imaging high-dimensional entanglement \cite{Edgar2012}, generalized Bell inequalities \cite{Dada2011} and the study of Einstein-Podolsky-Rosen non-localities \cite{Howell2004, Walborn2011}. However, these measurements used resource-intensive methods, such as sequential scanning or multiple standalone detectors. Early studies of entanglement with modern imagers used an electron-multiplying CCD (EMCCD) camera with an effective area of $201\times201$ pixels and frame readout-rate of $5Hz$ \cite{Edgar2012}. Albeit the EMCCD quantum efficiency was up to $90\%$, prolonged exposure time of about $1ms$, requires this device to operate at very small photon-rates to avoid multiple photons in the same frame. Furthermore, to achieve single-photon level sensitivity the EMCCD camera operated at a low temperature of $-85^{o}$C.\\

\noindent Further progression on quantum imaging with cameras was achieved using intensified CMOS and CCD cameras \cite{Unternahrer2016,Jost1998,Just2014,Fickler2013,Reichert2017,Reichert2017_2}. Flexible readout architectures allow $kHz$ continuous framing rates in CMOS cameras. Additionally, nano-second scale time resolution for single photons can be achieved by gating image intensifiers. For example, an intensified sCMOS camera was used to observe Hong-Ou-Mandel interference \cite{Jachura2015}, where the photons were collected on a $700\times22$ pixel area at a framing rate of $7kHz$. The photon acquisition statistics can also be enhanced by using multiple triggers during a single frame, so the camera integrates multiple photons within a single acquired image. This approach was employed in \cite{Fickler2013}, where an idler photon from an entangled pair was used to trigger an intensified CCD camera. Although many thousands of photons were imaged in a single frame of the camera, allowing the spatial characterization of the photon's angular momentum, the framing-rate was only $4Hz$.\\

\noindent This low throughput is a severe limitation to resolve spatial characterization of entanglement in real-time. Here we show how a development from the high-energy physics community, the intensified Tpx3Cam camera \cite{asi} can be converted into a quantum characterization device of photonic polarization entanglement. This setup allows for imaging and time-stamping of a continuous stream of entangled photons with an excellent spatial and temporal resolution ($55\times 55 ~\mathrm{\mu m^2}$, $1.5 ~\mathrm{ns}$), providing a high signal-to-background ratio. We emphasize that the Tpx3Cam readout is data-driven, with a high throughput of $\approx 10^7$ photons per second, which is several orders of magnitude higher than the conventional cameras discussed above.

\section{Experimental setup}
\begin{figure*}[h]
\centering
\includegraphics[width=1\columnwidth]{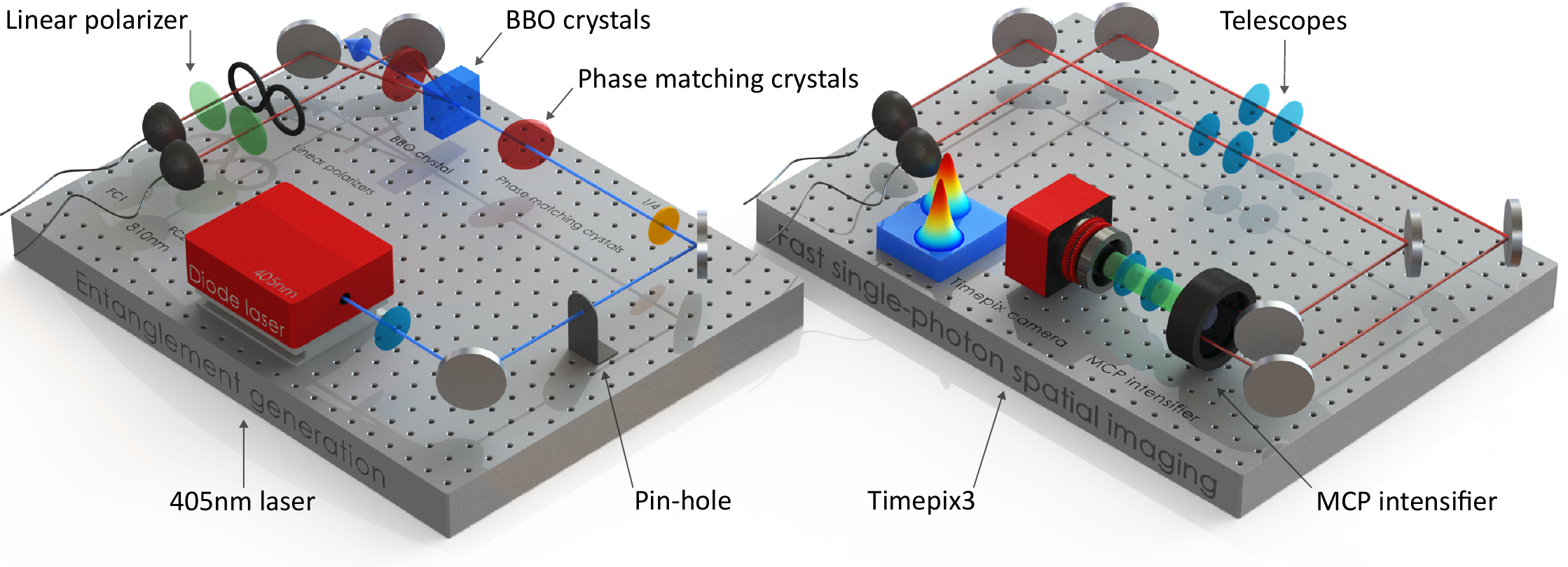}
\caption{\textbf{Experimental layout for entanglement generation (1; left) and characterization (2; right)}: 1) The source utilizes a blue pump laser diode tuned to a wavelength of $\lambda=405~\mathrm{nm}$ and a pair of Type I BBO crystals with optical axes perpendicular to one another to generate signal and idler photons entangled in polarization at a wavelength of $\lambda=810~\mathrm{nm}$. The entangled photons undergo individual transformations using polarizers to evaluate Bell's inequality parameters and are then fiber coupled. 2) Exiting from the fibers photons are mode-matched and detected by an image intensifier before registration with the Tpx3Cam camera.}
\label{fig:setup}
\end{figure*}

\noindent In this collaboration experiment between Stony Brook University and Brookhaven National Laboratory, we study the characterization of SPDC-based quantum polarization entanglement using fast $2D$ imaging with a Tpx3Cam. The experimental layout is shown in Figure \ref{fig:setup}.

\noindent \subsection{Entangled-photon source}
\noindent Our entanglement source (QuTools QuED \cite{Pomarico2011}) utilizes a blue pump laser diode tuned to wavelength $\lambda=405~\mathrm{nm}$, and a pair of Type I non-collinear BBO crystals with optical axes perpendicular to one another to generate signal and idler photons at wavelength $\lambda=810~\mathrm{nm}$. The first crystal optical axis and the pump beam define the vertical plane. Owing to Type-I phase matching, an incoming photon which is vertically polarized gets down-converted and produces two horizontally polarized photons in the first crystal, whereas a horizontally-polarized photon going through these crystals would get down-converted in the second crystal producing two vertical photons. An additional pair of birefringent crystals ensures maximum spatial overlap of the down-converted photons by pre- and post- compensating for differences in effective optical path lengths of signal and idler. The produced state has the form:

$$\ket{\phi^\pm}=\frac{(\ket{HH}\pm\ket{VV})}{\sqrt{2}}$$.

\noindent Signal and idler photons are spatially separated and collected using single-mode, polarization non-maintaining fibers, with a linear polarizer, used for projective measurements, before each coupler.

\noindent \subsection{Intensified fast-camera: Tpx3Cam}
\noindent An intensified camera, Tpx3Cam \cite{asi}, achieves imaging with single-photon sensitivity, which allows time-stamping of incident photons with $1.5~\mathrm{ns}$ time granulation when coupled to an image intensifier. The Tpx3Cam consists of a light-sensitive silicon sensor bump-bonded to Timepix3, a time-stamping readout chip \cite{timepix3}. The sensor-chip assembly has $256\times256$ pixels of $55\times55$ $\mathrm{\mu m^2}$ each. The silicon sensor in the camera has a thin entrance window with an anti-reflective coating providing excellent quantum efficiency \cite{sensor}. The sensor is optimized for emission spectrum of the $P47$ scintillator \cite{sensor2}. The non-intensified version of Tpx3Cam has been used before for the velocity mapped ion imaging \cite{tpx3cam} while the intensified version of its predecessor, TimepixCam, has been used for fluorescent lifetime imaging, which required single photon sensitivity, similar to this application \cite{intensifiedtimepixcam}.\\

\noindent  The Timepix3 processing electronics in each pixel records the time-of-arrival (TOA) of hits that cross a preset threshold and stores it as a time-code in a memory inside the pixel. The time-over-threshold (TOT) is also recorded serving as an estimate of the light flux seen by the pixel. The individual pixel dead time is of the order of $1~\mathrm{\mu s}$. The readout is data-driven, and only the pixels with signals above the threshold are recorded. The camera can operate continuously and does not require a trigger as the pixels transfer the data asynchronously within microseconds after being hit. The maximum camera throughput is $80 ~\mathrm{Mpix/s}$ \cite{asi,spidr}.\\

\noindent  The intensifier in front of the camera is an off-the-shelf vacuum device \cite{intensifier} with a photocathode followed by a chevron micro-channel plate (MCP) and fast $P47$ scintillator with a signal rise time of $\sim 7 ~\mathrm{ns}$ \cite{p47}. Photons are first converted to photoelectrons in the photocathode and then amplified in the MCP before producing a flash of light in the scintillator. The $18~\mathrm{mm}$ diameter scintillator screen is projected on to the $14\times14 ~\mathrm{mm^2}$ sensor with a relay lens with no magnification \cite{cricket}. The photocathode used for the experiments had a quantum efficiency attaining $\approx 18\%$ for the used wavelength of $810~\mathrm{nm}$.\\

\noindent  The camera was calibrated to equalize the response of all pixels by adjusting the individual pixel thresholds. After this procedure, the effective threshold to fast light flashes from the intensifier is 700-800 photons per pixel depending on the wavelength. A small ($\approx0.1\%$) number of hot pixels was masked to prevent logging large rates of meaningless data to the disk.

\section{Experimental results}
\noindent \subsection{Benchmarking: entanglement characterization.}
\noindent Our procedure starts by evaluating the entangled state produced by the source. We assume the state to be in a superposition of two Bell-states of the form:
\begin{eqnarray}
\ket{\psi}=\cos\theta\ket{\phi^+}+e^{i\delta}\sin\theta\ket{\phi^-} = \frac{(\cos\theta+e^{i\delta}\sin\theta)}{\sqrt{2}}\ket{HH} + \frac{(\cos\theta-e^{i\delta}\sin\theta)}{\sqrt{2}}\ket{VV}
\end{eqnarray}


\noindent Using a density matrix $\rho=\ketbra{\psi}{\psi}$, after projection of the two photons by polarizers with angles $\alpha$ and $\beta$, we obtain an expectation value for the measurements of coincidences:
\begin{eqnarray}
P_{VV}(\alpha,\beta)=\text{Tr}\{\rho\hat{M}_{\alpha\beta}\}=c_0+c_1\cos2\beta+c_2\sin2\beta
\end{eqnarray}

\noindent Here, the operator $\hat{M}_{\alpha\beta}=\ketbra{V_\alpha V_\beta}{V_\alpha V_\beta}$ denotes the projection onto a vertical polarization state. In the basis of BBO crystal we have:
\begin{eqnarray}
\ket{V_\alpha V_\beta}=\sin\alpha\sin\beta\ket{HH}-\sin\alpha\cos\beta\ket{HV}-\cos\alpha\sin\beta\ket{VH}+\cos\alpha\cos\beta\ket{VV}
\end{eqnarray}
\noindent with $c_0=\frac{1-\sin2\theta\cos{\delta}\cos2\alpha}{4},c_1=\frac{\cos(2\alpha)-\cos\delta\sin2\theta}{4}$ and $c_2=\frac{\cos2\theta\sin2\alpha}{4}$.\\

\noindent The incoming photon pairs from BBO crystal and background are denoted as $N_0$ and $N_d$ respectively. Then the total coincidence can be fitted as the equation:
\begin{eqnarray}
N(\alpha,\beta)=N_0P_{VV}(\alpha,\beta)+N_d = C_0+C_1\cos(2\beta)+C_2\sin(2\beta)
\end{eqnarray}
\noindent where $ C_0=-\frac{N_0\cos\delta\sin{2\theta}}{4}\cos{2\alpha}+\frac{N_0+4N_d}{4}, C_1=\frac{N_0}{4}\cos{2\alpha}-\frac{N_0\cos\delta\sin{2\theta}}{4}$ and $ C_2=\frac{N_0\cos{2\theta}}{4}\sin{2\alpha}$.\\

\noindent Experimentally, we evaluate the rate of coincidences using two single-photon counting modules, as a dependence of the polarization angles $\alpha$ and $\beta$, which are set by rotating two polarizers (cf. Fig. 1). The coincidence data for different settings of the polarizers and the respective fitting curves are shown in Fig. 2, where we see the oscillation predicted by the simple theory described above. We numerically fit the parameters $N_0$, $\theta$, $\delta$ and $N_d$ to the experimental data, obtaining the following results:
$N_0\pm \Delta N_0 =47640\pm2800$, $N_d \pm\Delta N_d=380\pm830$, $\theta \pm \Delta\theta =-0.15\pm0.10$ and $\delta \pm \Delta\delta =2.10\pm0.48$. Hence, the produced entangled state is: $\ket{\psi}=0.989\ket{\phi^+}+(0.076-0.130 i)\ket{\phi^-}$.

\begin{figure*}[h]
\includegraphics[width=1.0\columnwidth]{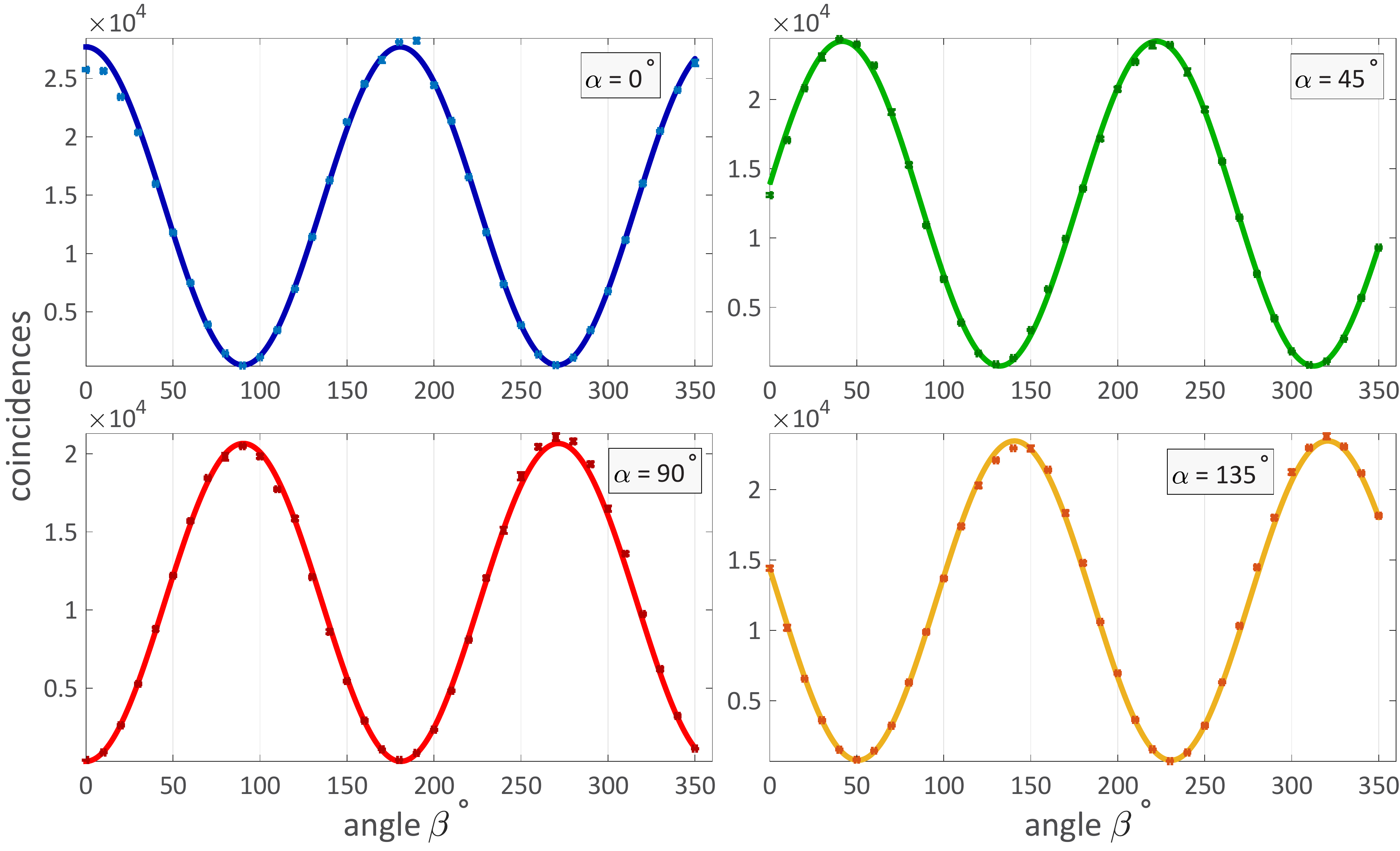}
\caption{\textbf{Coincidence for all four linear polarizer angles} $\alpha=0^\circ$ (blue), $45^\circ$ (green), $90^\circ$(red), $135^\circ$(orange) using quTools \cite{qutools}. Polarizer angle $\beta$ was varied over full 360 $^\circ$ at a step of 10$^\circ$ for each of four $\alpha$'s. Five data points were taken and averaged at each polarizer combination. The uncertainty of polarizer angles is 1$^\circ$. Curves are fitted with sine functions predicted from pure state model as discussed in entanglement characterization. The pure quantum state is fitted to be $\ket{\psi}=0.989\ket{\phi^+}+(0.076-0.130 i)\ket{\phi^-}$. Notice that the colors are chosen to be consistent with the following measurement in Fig. \ref{fig:s-value}. }
\label{fig:C01}
\end{figure*}

\noindent \subsection{Benchmarking: Bell inequality violation using SPCM.}
\noindent Our next step is to calculate the Clauser-Horne-Shimony-Holt (CHSH) inequality violation \cite{Aspect1981,Aspect1982} using SPCM (Single Photon Counting Module). The inequality can be written as
\begin{eqnarray}\label{chsh}
S=E(\alpha,\beta)+E(\alpha',\beta)-E(\alpha,\beta')+E(\alpha',\beta')\le 2
\end{eqnarray}where
$E(\alpha,\beta)=\frac{N_{VV}(\alpha,\beta)+N_{HH}(\alpha,\beta)-N_{VH}(\alpha,\beta)-N_{HV}(\alpha,\beta)}{N_{VV}(\alpha,\beta)+N_{HH}(\alpha,\beta)+N_{VH}(\alpha,\beta)+N_{HV}(\alpha,\beta)}$ from the fitted curves in Fig. 2. We obtain the values shown in Table \ref{tab:table1}.\\

\begin{table}
\caption{\label{tab:table1}The number of coincidences used to calculate the S-value.}
\begin{center}
\begin{tabular}{ |c|c|c|c|c|c| }
\hline
$(\alpha,\beta)^\circ$ & $N_{VV}$ & $N_{HV}$ & $N_{VH}$ & $N_{HH}$ & $E(\alpha,\beta)$ \\ \hline
(0,22.5)&17656 & 4393 & 3344& 23767 &0.685243\\ \hline
(0,67.5)&3344 & 23767 & 17656& 4393 &-0.685243 \\ \hline
(45,22.5)&19064 & 2984 & 5516& 21596 &0.654174\\ \hline
(45,67.5)&21596 & 5516 & 2984& 19064 &0.654174 \\\hline
\end{tabular}
\end{center}
\end{table}

\noindent Using these values, we calculate the $S$ parameter 
$S=2.679 \pm 0.007 >2$,
clearly showing the violation of the CHSH inequality. The uncertainty is calculated using $\Delta S=\sqrt{\sum_{\alpha,\beta}{\Delta E(\alpha,\beta)^2}}$ and
\begin{eqnarray}
 \Delta E&=&\frac{2[N_{VV}(\alpha,\beta)+N_{HH}(\alpha,\beta)][N_{HV}(\alpha,\beta)+N_{HV}(\alpha,\beta)]}{\left(N_{VV}(\alpha,\beta)+N_{HH}(\alpha,\beta)+N_{HV}(\alpha,\beta)+N_{HV}(\alpha,\beta)\right)^2}\nonumber\\
 &\times&\sqrt{\frac{1}{N_{VV}(\alpha,\beta)+N_{HH}(\alpha,\beta)}+\frac{1}{N_{HV}(\alpha,\beta)+N_{VH}(\alpha,\beta)}}
\end{eqnarray}

\section{Entanglement characterization with Tpx3Cam}
\noindent Having set a benchmark for the measurements, we now proceed to characterize the entanglement source using  the Tpx3Cam. Instead of being measured in the single-photon detectors, we now send the entangled pairs to another experimental setup where they are converted to photoelectrons, amplified in different regions of the intensifier and sub-sequentially time-stamped in the fast camera.\\

\begin{figure*}
    \centering
    \begin{subfigure}{.45\textwidth}
         \centering
        \caption{}
        \includegraphics[width=0.98\linewidth]{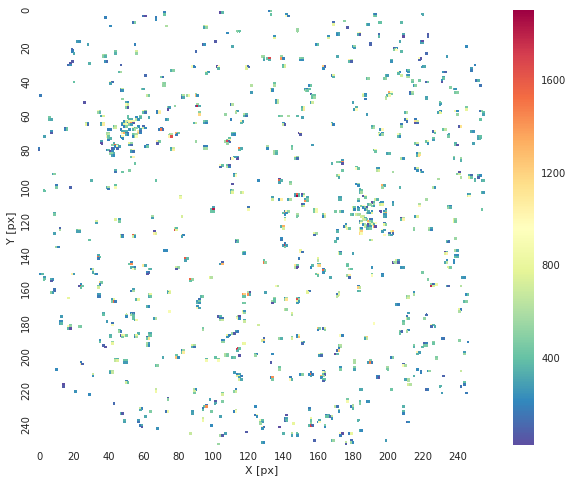}

    \end{subfigure}
\begin{subfigure}{.45\textwidth}
 \centering
 \caption{}
 \includegraphics[width=0.98\linewidth]{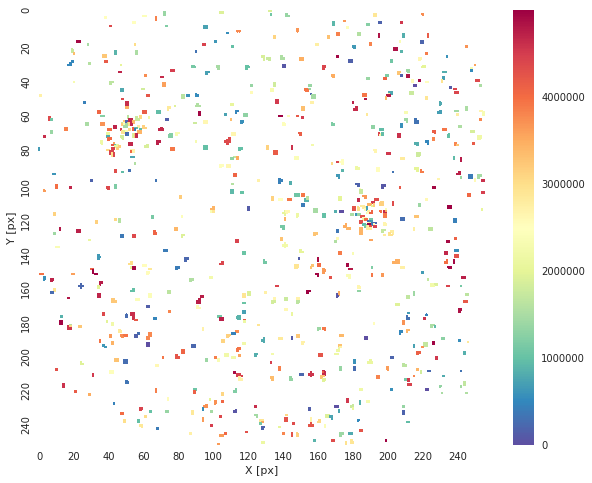}

\end{subfigure}

\begin{subfigure}{.45\textwidth}
 \centering
 \caption{}
\includegraphics[width=0.98\linewidth,height=6.2cm, keepaspectratio]{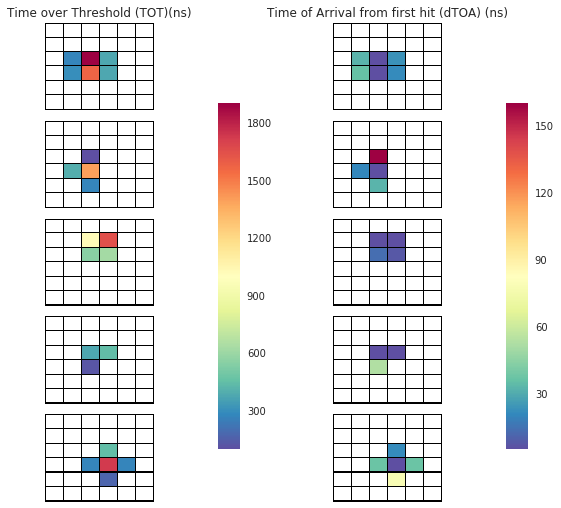}
\end{subfigure}
\begin{subfigure}{.45\textwidth}
 \centering
  \caption{}
 \includegraphics[width=0.98\linewidth]{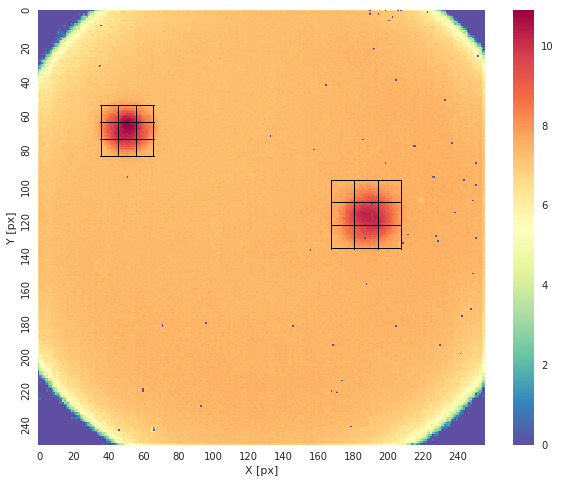}

\end{subfigure}

\caption{a) Single photon hits recorded by the camera in a time slice of $5 ms$. The color encodes the time-over-threshold (TOT) in ns. b) same single photon hits recorded by the camera in a time slice of $5 ms$. However this time, the color encodes the time-of-arrival (TOA) in ns. c) Examples of zoomed-in photon hits. The left column shows TOT in ns; the right column shows relative TOA from the first hit pixel in ns. d) $2D$ occupancy map of the sensor ($256\times256$ pixels) showing the photon hits for the full statistics of a five-minute run. The color encodes the number of times a particular pixel was hit in log scale.}
\label{fig:photons1}
\end{figure*}

\noindent The photon pairs were recorded continuously by the camera for a given period of time, $5$ minutes for each combination of polarizations. Figure \ref{fig:photons1} shows single photon hits recorded by the camera in a time slice of 5 ms, examples of the hits registered in the camera and the pixel occupancy map. Note that photons can be recorded by the same pixel multiple times during the $5 ms$ time period as the pixel deadtime is only $1\mu$s. If this is the case the latest pixel information is shown Figures \ref{fig:photons1}a) and \ref{fig:photons1}b). The two fiber-modes are clearly visible, corresponding to the areas of highest occupancy. The intensity distributions in the fibers follow the Gaussian modes as expected. The rate of photons in these regions was $\approx 30~\mathrm{kHz}$, determined primarily by the output rate of the photon source at the fiber end (typically $\approx 150 ~\mathrm{kHz}$ ) and the intensifier quantum efficiency.\\

\noindent The background, uniformly distributed over the photocathode surface in the occupancy map in Figure \ref{fig:photons1}d), is caused by spurious dark counts from the photocathode. This rate is $\approx 50$ times smaller than the measured single photon rate and could be further reduced by cooling the intensifier, which in our measurements was operated at room temperature. We also note that the background photo-electrons arrive at random times and thus will be suppressed by requiring coincidence between the two photons, as shown below.

\noindent \subsection{Data processing}
\noindent To perform a Bell's inequality measurement using data from the fast camera, we gathered $72$ five-minute long datasets corresponding to different settings of the polarizers. The raw data is processed following several steps: \textit{i}) time-ordering of the hit pixels, \textit{ii}) identification of the pixel clusters corresponding to the single photon hits, \textit{iii}) centroiding of the pixel clusters, \textit{iv}) TOT corrections to improve the time resolution, \textit{v}) calculating the number of coincidences and \textit{vi}) Bell analysis.\\

\noindent \textbf{I. Time-ordering:} Tpx3Cam reads out the hit pixels asynchronously, which might alter the time order, especially at high rates. Thus, the first step of the data processing, is to time-order them to prepare the data for the cluster finding.\\

\noindent \textbf{II. Clustering:} Clusters are groups of pixels adjacent to each other and within a preset time window. We used a recursive algorithm to look for the clusters: for a pixel, a $1~\mathrm{\mu s}$ time window is applied to select other pixels close in space and time to the first one. Each pixel in a cluster should have a neighboring pixel separated not more than 300ns. The algorithm then chooses another pixel, not contained in a cluster, shifting the time window and starting the process anew.
\\

\noindent \textbf{III. Centroiding:} A photon hit in the camera comprises on average $\approx 4$ pixels with measured TOA and TOT, which allows applying a centroiding algorithm. The TOT information is used as a weighting factor, helping to define the geometrical center of the cluster, yielding an estimate of the coordinates $x$, $y$ of the incoming single photon. The arrival time of the photon is estimated by using TOA of the pixel with the largest TOT in the cluster. The above TOA is corrected as described below.\\

\noindent \textbf{IV. TOT correction:} Photons in the entangled pairs are simultaneous. Therefore, they will have the same time-stamps, within the time resolution. Precise timing is a powerful handle to reduce the random background, thus improving the camera time resolution and the signal-to-background ratio. For this, the timing information must be corrected to account for the so-called time walk. In the Timepix3 front-end electronics, within each pixel, the discriminator has a constant threshold, so larger signals cross the threshold earlier producing smaller TOA and larger TOT values. The correlation of ToA and ToT allows to calibrate the constant threshold effect and, hence, to improve the time resolution.\\

\begin{figure*}[h]
\centering
\includegraphics[width=1.0\columnwidth]{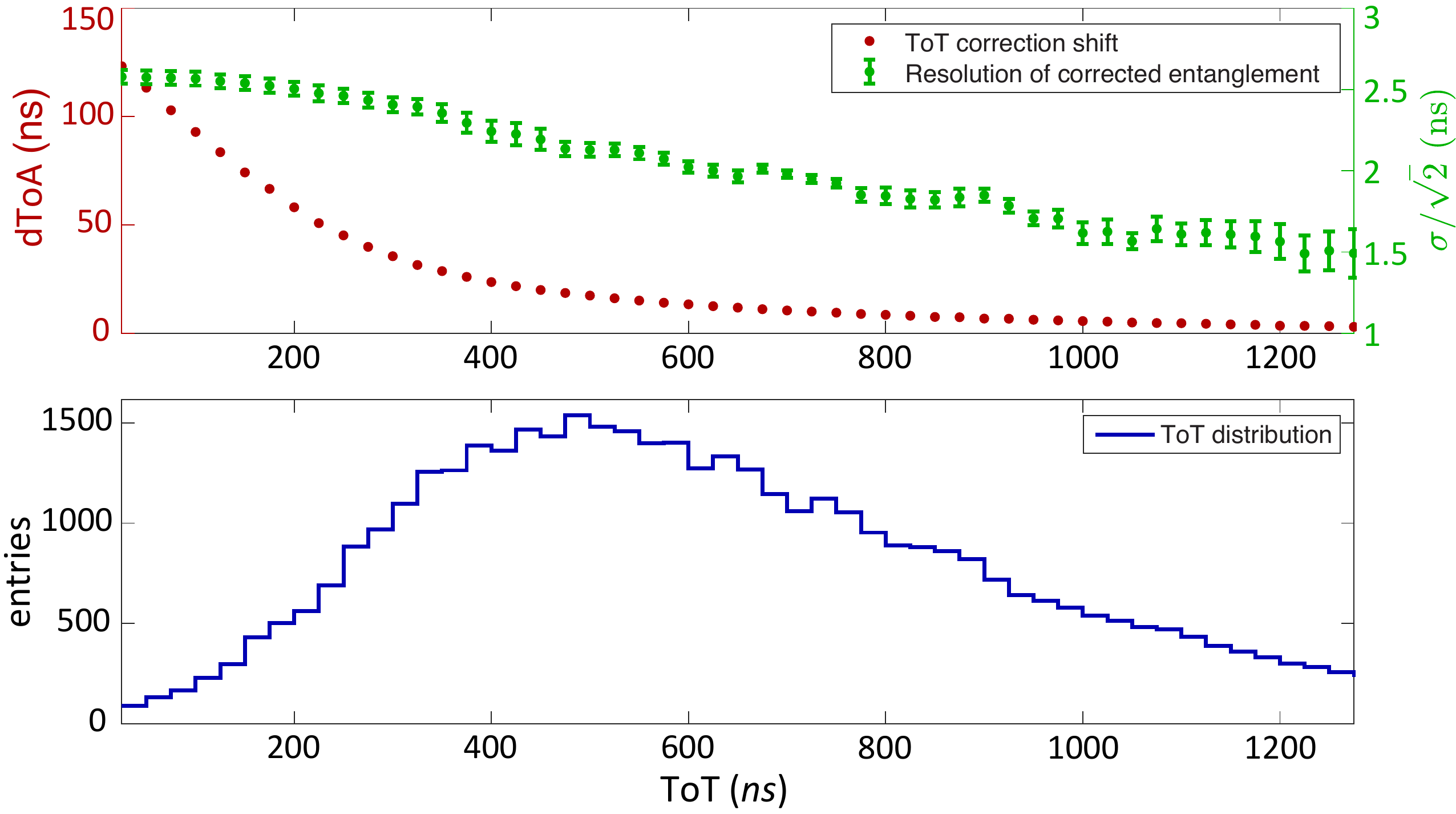}
\caption{Time correction and time resolution as a function of TOT (top) and TOT distribution (bottom). In the top figure, the red dots show the TOA shift due to the TOT correction.
The green dots show the time resolution as function of TOT.}
\label{fig:TOT}
\end{figure*}

\noindent Typically, the correction requires a time reference, for example from a laser, to determine the shift, as in the ion imaging experiments \cite{tpx3cam}. However, in these experiments, the entangled pairs are generated continuously, so a different procedure, which does not rely on an external time reference had to be developed. This was done as follows: the zero reference time of each cluster was defined as the TOA of the earliest pixel in the cluster, TOA$_\mathrm{centroid}$, so a time difference (dTOA = TOA - TOA$_\mathrm{centroid}$) can be calculated for each pixel in the cluster and associated with the pixel TOT value. Using a range of large TOT values where the dTOA was stable (typically for TOT$_\mathrm{centroid}$ greater than 1500ns there is no change in dTOA) as a global time reference, a lookup table of dTOA shifts for different TOT values was obtained, shown in the top graph of Figure \ref{fig:TOT}. This procedure reduces the time difference between entries within a given cluster from $\sim 100~\mathrm{ns}$ to a few ns correcting the time walk for individual pixels.\\

\noindent The time resolution after the TOT correction is shown in the top graph of Figure \ref{fig:TOT} as a function of TOT. The time resolution is determined from the distribution of time difference between two entangled photons. The distribution is fit with a Gaussian function, and the resolution is defined as the sigma of the fit divided by $\sqrt{2}$. This accounts for the fact that variance of the time difference between two photons is larger than the resolution per photon. Therefore the cited time resolution is per photon assuming equal resolution for each of two photons. To determine the dependence of the resolution on TOT we required that both photons have TOT greater than a certain value on the TOT axis.
The bottom graph shows the distribution of the TOT values.\\

\begin{figure*}[h]
\centering
\includegraphics[width=1\columnwidth]{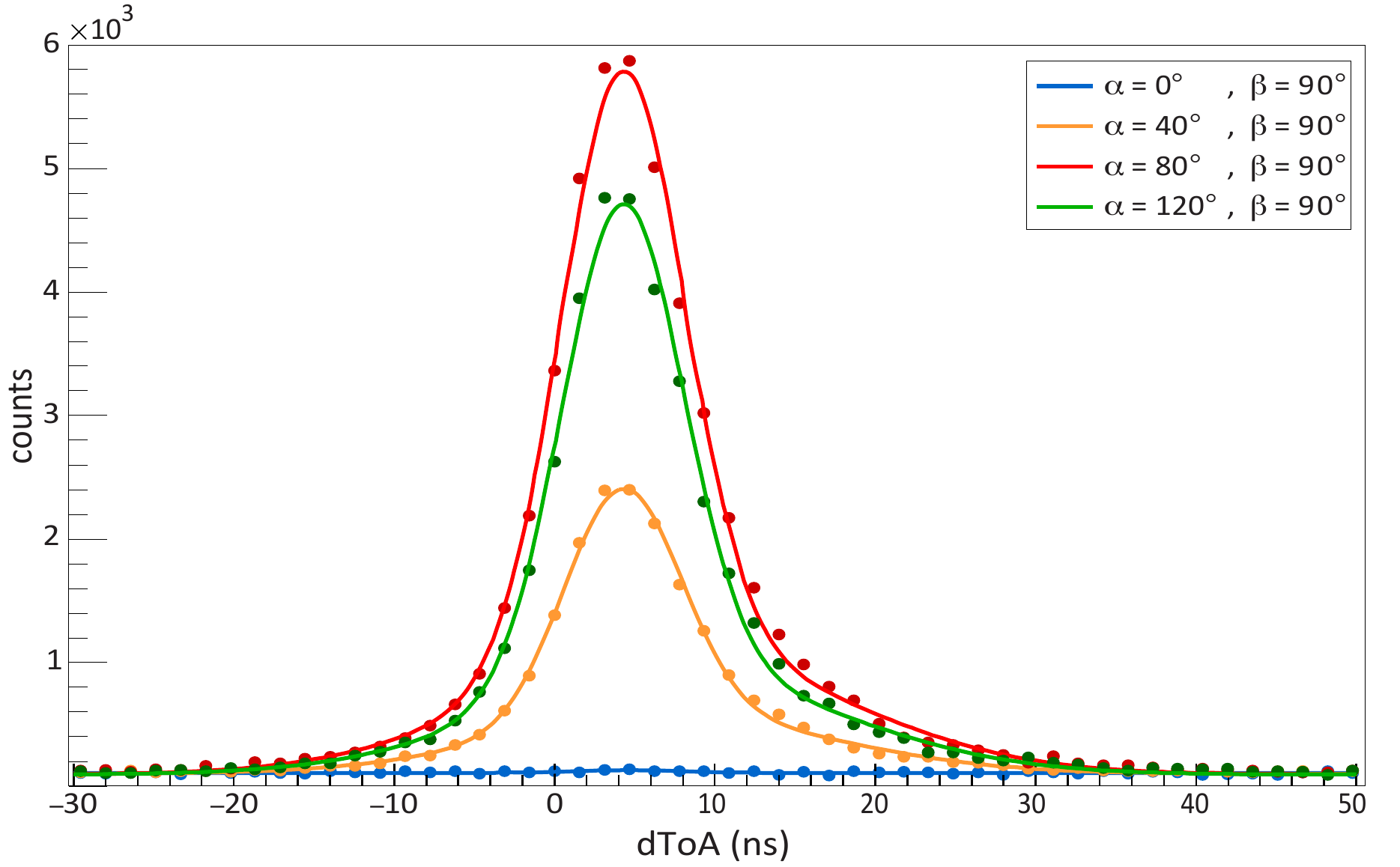}
\caption{Distribution of time difference between two photons in different fibers for selected pairs of polarization settings combinations: $\alpha=0^\circ$ (blue), $\alpha=40^\circ$ (orange), $\alpha=80^\circ$ (red) and $\alpha=120^\circ$ (green) with $\beta=90^\circ$. The amplitudes for different polarization combinations are determined by the entangled state projection. Flat background is the result of uncorrelated photon background. The total number of coincidences is calculated by integration over the Gaussian curves.}
\label{fig:dtime}
\end{figure*}

\noindent \textbf{V. Time coincidences:} to identify pairs of simultaneous photons we selected areas of the sensor corresponding to regions illuminated by the fibers. The corresponding square areas were $42\times42$ pixels for the right fiber and $30\times30$ pixels for the left fiber as they are shown in Figure \ref{fig:photons1}d). Then, for each photon detected in one region, we looked for its associated pair at the closest time in the second region. The time difference distribution for these detected pairs is shown in Figure \ref{fig:dtime} for several settings of the polarizers, as defined in the Bell measurement above. The prominent peaks correspond to the pairs of entangled photons, while the small flat background corresponds to random coincidences of uncorrelated photons. Due to the finite quantum efficiency and other losses, not each photon from the source will have a detected synchronous partner in the other fiber. In this case, it will be paired with a random photon, either from the photon source (more likely) or from the spurious photocathode counts, giving rise to the flat background.

\noindent Each distribution was fit to a function consisting of two Gaussians and a constant accounting for the flat background of random coincidences. The number of coincidences was estimated as the area under the Gaussian functions. The dependence of the number of coincidences on the polarizer settings indicates that the operation of the camera detection setup closely resembles the SPCM operation, despite the use of an entirely different registration scheme for single photons.

\noindent \subsection{Bell's inequality violation with a fast camera.}
\label{sec:results}
\begin{figure*}[h]
\centering
\includegraphics[width=1\columnwidth]{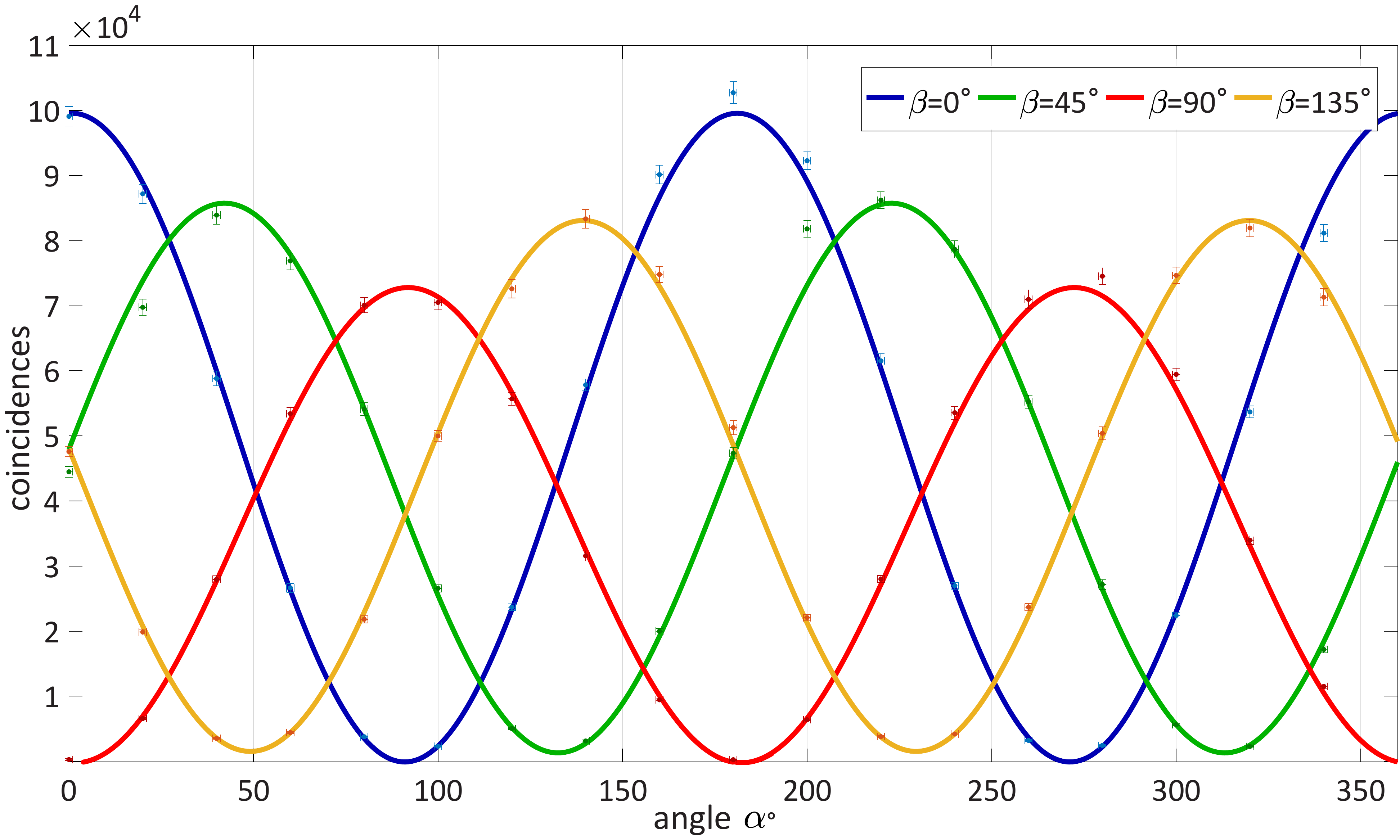}
\caption{\textbf{Coincidences as a function of polarization:} showing the dependence of the signal amplitude (number of coincidences) for different settings of a Clauser-Horne-Shimony-Holt type Bell inequality test. We use the same color regulation as Figure \ref{fig:C01}, with the fix polarization $\beta=0^\circ$ (blue), $\beta=45^\circ$(green), $\beta=90^\circ$(red) and $\beta=135^\circ$(orange). The uncertainty in polarization is  1$^\circ$. Fitting parameters can be found in Table \ref{tab:table4}. }
\label{fig:s-value}
\end{figure*}

\noindent Our next step is to study the dependence of the coincidence measurements on the polarization projective measurements of the two-photon state. In the measurements, one polarizer was varied in $20^\circ$ steps for four fixed values of the other polarizer: $0^\circ$, $45^\circ$, $90^\circ$ and $135^\circ$. The dependence of the number of coincidences versus the polarizer angle is shown in Figure \ref{fig:s-value}. The data points were fitted with a sine function with the period, phase, amplitude and offset as free parameters. The fit results are shown in Table \ref{tab:table4}.\\

\begin{table}
\caption{\label{tab:table4}The parameters of the sin functions used to fit the polarization function. The function $A \sin(\frac{2\pi}{T}(x+\phi))+D$ is used to fit the values. Here amplitude is defined as $A$, $T$ is period, phase shift is $\phi$,  and offset is $D$.}
\begin{tabular}{|c|c|c|c|c| }
\hline
Polarization $\beta$ & Amplitude $A$ &\mbox{Period $T$ [deg]}&\mbox{Phase shift $\phi$} [deg] &\mbox{Offset $D$}\\
\hline
{\color{blue} \textbf{0}} & $1.000  $ & $180.2\pm0.2$ & $ -44.0 \pm0.2$ & $0.999\pm0.005$\\ \hline
{\color{olivegreen} \textbf{45}} & $0.851\pm0.005$ & $180.8\pm0.2$ & $ -3.8\pm0.2$ & $0.877\pm0.004$ \\ \hline
{\color{red} \textbf{90}} & $0.735\pm0.004$ & $180.5\pm0.2$ & $ 46.5\pm0.2$ & $0.732\pm0.004$ \\ \hline
{\color{orange} \textbf{135}} & $0.822\pm0.005$ & $180.2\pm0.2$ & $ 94.2\pm0.2$ & $0.853\pm0.004$ \\ \hline
\end{tabular}
\label{table:1}
\end{table}

\noindent With these experimental parameters, we follow the procedure outlined in the benchmarking section, to obtain a Bell-state inequality S-value with the Tpx3Cam setup. The obtained value is $2.78 \pm 0.02$, well above the classical limit of 2, and closer to the Tsirelson Bound of $2\sqrt{2}$ ($2.82$) than the SPCM measurements. We attribute this improvement to the better rejection of random background enabled by the fast camera.

\noindent \subsection{Position dependent Bell analysis.}
\begin{figure*}[h]
\centering \includegraphics[width=1\columnwidth]{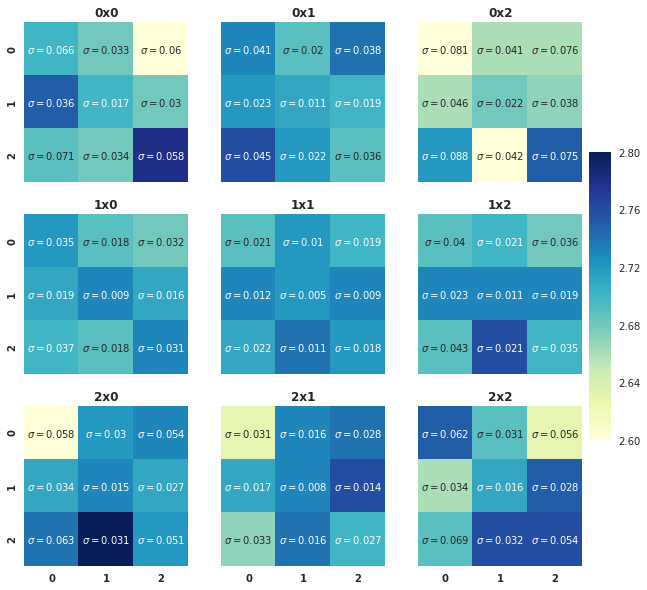}
\caption{Table of S-values for subarea matrices. In this configuration the two areas on the fast camera that are illuminated by photons from the fibers are divided into subareas, forming two $3\times 3$ matrices. The coincidence thus decomposes into that of $81$ possible pairs of a combination of subareas. Using these coincidences, we calculated the CHSH inequality violation and plotted the resulted S values in the form of $81$ blocks in nine $3\times3$ matrices. The S-values are color coded with the corresponding uncertainty shown in the center of the box. This gives an intuitive illustration of the spatial distribution of entangled photon pairs. The black digits above each matrix give the position of the photon in the first fiber: $0\times 0$ corresponds to the top left corner, $0\times 2$ to the top right corner, $1\times 1$ to the center.}
\label{fig:matrix}
\end{figure*}

\noindent One of the clear advantages of using high-speed cameras for quantum state characterization is the capacity to analyze multiple processes simultaneously. In our last experiment, we probe the capacity of the fast camera to analyze $81$ entangled pairs in parallel. To simulate the latter, we divided the areas illuminated by the fiber's output into nine subareas, forming a $3\times3$ matrix as shown in the two fiber regions in Figure 3. We then analyze each pair-wise combination ($81$ total combinations) independently and reproduce the Bell analysis presented above for each of them. In order to accumulate enough statistics for these spatially resolved measurements, we took one-hour-long extended datasets, corresponding to the $16$ combinations of the polarizers settings, which are needed to calculate the Bell's inequality violation. The total number of recorded photons was considerable, about $10^9$, and required efforts to implement parallel processing of the data. The data analysis was performed within the modular scientific software framework $root$ developed at CERN \cite{root}. Figure \ref{fig:matrix} shows the results of the parallel evaluation of $81$ Bell's inequalities. Each box represents a specific spatial combination of subareas. The corresponding S-value is color-coded with the corresponding uncertainty shown in the center of the box. The results show that the S-value is uniform within the experimental errors with no position dependence visible.

\section{Discussion and outlook}

\noindent We have demonstrated that the spatial characterization of photonic entanglement can be performed employing the intensified Tpx3Cam camera. The camera can simultaneously time-stamp multiple single optical photons with nanosecond timing resolution while also capturing their spatial information. The S-value results confirm that the fast camera spatial characterization of quantum-states in parallel is a viable alternative to be used in scaled up quantum systems.\\

\noindent The photon rate in these experiments was limited by the photon source to about 100 kHz. This is a factor of 100 lower than the maximum rate allowed by the camera, which should be capable of working with much brighter sources of entangled photons. Another specification of the fast camera, the quantum efficiency (QE) of the image intensifier, is another critical parameter for the efficient detection of entangled photons. New photocathodes based on GaAs offer QE of $\sim 35\%$ in the same wavelength range as used for these studies \cite{newintensifier}.\\

\noindent New imaging technologies based on monolithic silicon devices, such as SPADs (Single Photon Avalanche Devices) are rapidly improving and could become competitive soon.  Since the devices have internal amplification, the image intensifier is not required, which is a considerable simplification. In addition, SPADs could have better time resolution and, potentially, higher QE, compared to the intensified cameras. The SPAD imagers started to appear on the market, and first applications for QIS have been reported \cite{spad}. Currently, the main limitation of the devices is the high dark count rate in the tens of MHz/cm$^{2}$ range, which may saturate the readout and lead to low signal-to-background ratio. Another difficulty is the integration of the photon sensing SPAD pixels and complex readout electronics in a monolithic device, which has many technical challenges. Also, in a SPAD, a single photon fires only one pixel producing a standard pulse so no centroiding is possible and, therefore, it is also impossible to distinguish a noise hit from useful signal.\\

\noindent From the QIS perspective, we have showcased the possibility of parallel processing of tens of entangled states in parallel by analyzing independent combinations of subareas illuminated by the two fibers, which is an unprecedented capability for quantum information processing. As all pixels of the Tpx3Cm sensor act independently of each other, the dimensionality of this processing can be scaled up many-fold, for example, employing the same camera setup with brighter photon sources and/or multiple photon beams. We estimate that Tpx3Cam can successfully process at least $10\times10=100$ photon beams, each with a photon rate similar to the one used in these experiments. This technique may become a crucial tool for the real-time characterization of the performance for large entanglement-based quantum networks or circuits.\\

\noindent The camera also has the ability to count the number of spontaneous photons in the same fiber, given sufficient spatial separation. This offers the possibility of discerning an event with more than one photon pairs, an effect of the statistical distribution of the number of photons at the output of the SPDC process. This information was not used in the present analysis.\\

\noindent We also envision that our characterization setup can prove effective in other quantum information processing tasks, such as Hong-Ou-Mandel interference \cite{Jachura2015} and the characterization of entanglement encoded in orbital angular momentum modes \cite{Fickler2012}. Furthermore, it is well suited for the real-time benchmarking of quantum memories using OAM states \cite{Nicolas2014, Ding2015}, and for the parallel processing of the information in many memories systems \cite{Parniak2017}. Lastly, it could also find a niche as a feedback tool in the positioning of long-distance free-space quantum communication channels forming intra-city quantum cryptographic networks \cite{Steinlechner2017}.


\section{Acknowledgements}
\noindent The authors kindly thank Andrea Londono and Julia Tsybysheva for their technical assistance with the measurements; Heinz Graafsma for providing the Timepix3 ASIC used in these experiments; Martin van Beuzekom, Bram Bouwens and Erik Maddox for their assistance with the fast camera; Thomas Tsang for assistance with the image intensifier. The Stony Brook team acknowledges the support from the National Science Foundation (grants PHY-1404398 and PHY-1707919) and the Simons Foundation (grant SBF241180). Furthermore, this work was supported by Czech Ministry of Education, Youth and Sports (grant LM2015054) and by the LDRD grant 18-051 of Brookhaven National Laboratory. A.H. acknowledges support under the Science Undergraduate Laboratory Internships Program (SULI) by the U.S. Department of Energy.




\end{document}